\documentclass[12pt]{article}
\usepackage{epsf}

\newcommand{\be}{\begin{equation}}
\newcommand{\dl}{\delta}
\newcommand{\ee}{\end{equation}}
\newcommand{\lll}{\langle}
\newcommand{\rrr}{\rangle}
\title{\bf QCD Field Strength
Correlator at the One-Loop Order}
\author{V.I.Shevchenko\thanks{e-mail: shevchenko@vxitep.itep.ru} \\
\it Institute for Theoretical and Experimental Physics \\
\it 117218, B.Cheremushkinskaya 25, Moscow, Russia}
\date{}
\begin{document}
\maketitle
\vspace{1cm}
{\centerline {\bf Abstract}}

The leading perturbative contributions into the two-point gauge-invariant correlator
$Tr\langle gF_{\mu\nu}(x)U(x,0)U(0,y)gF_{\rho\sigma}(y)U(y,0)U(0,x)\rangle$
are calculated at the one-loop order. It is shown, that nonlocal
condensate $Tr \langle {\alpha}_s F_{\mu\nu}F_{\mu\nu}\rangle $ is nonzero
at this order. The relation with the renormalization properties
of Wilson loops is discussed.

\newpage

 Our present theoretical understanding of strong interactions is
based on quantum chromodynamics (QCD).
It is well known, that the gauge nature of this theory leads
to a wealth of its properties as well as to the great difficulties
in the description of its dynamics.
One source of these difficulties lies in the fact,
that despite all observables one can calculate in the theory must
be gauge invariant, the calculations actually involve noninvariant
quantities like propagators at intermediate steps.
The situation is studied well for the covariant gauges, at the same time
for the noncovariant ones a lot of
work still has to be done (see {\cite{leup}} and references therein).
There is also a kind of temptation to consider one set of gauge conditions
as "more physical" than another one supported by real simplifications
arising then one chooses the "most adequate gauge" for the given problem
(despite all physical observables are of course gauge independent). 

Therefore it is very useful if possible to reformulate the theory
in terms of gauge invariant quantities from the beginning.
The latter are usually taken as Wilson loop
functionals:
\be
W(C) = \lll Tr\>Pexp\>(ig\int\limits_{C} A_{\mu} dx^{\mu})\rrr
\label{w}
\ee
The ordinary brackets denote usual average:
$$
\lll O[A(x)]\rrr = \int DA(x) O[A(x)] exp\>\left(-\frac{1}{4} \int d^4x F^{a}_{\mu\nu}
F^{a}_{\mu\nu}\right)
$$
where field strength tensor is $ F^a_{\mu\nu} =
{\partial}_{\mu}A^a_{\nu} -
{\partial}_{\nu}A^a_{\mu} + g f^{abc} A^{b}_{\mu} A^{c}_{\nu}$

Unfortunately, the loop approach \cite{mm} has two undesirable features.
First of all, the original equations for loop
variables are written for nonrenormalized
quantities and complete renormalization program for them seems to be very
difficult to proceed. Second, the loop equations are
complicated since they defined not in the familiar coordinate
space but in the loop space -- subtle mathematical object itself.

All that makes attempts to look for an alternative set of gauge
invariant quantities attractive.
Recently a lot of work has been done (see \cite{sss}
and references therein) in order
to represent QCD on a way, where
dynamical degrees of freedom are gauge invariant cumulants of
the following type:
\be
D_{n} = \lll\lll\prod\limits_{i=1}^{n} G(x_i, x_0)\rrr\rrr
\label{cum}
\ee
$$
G(x_i, x_0) = U(x_0, x_i) gF_{{\mu}_i{\nu}_i}(x_i) U(x_i, x_0)
$$
where nonabelian phase factors are 
$$
U(x,y)=Pexp(-ig \int\limits_x^y A_{\mu}(z) dz_{\mu}) = 
$$
\be
= 1 - ig\int\limits_x^y A_{\mu}(z) dz_{\mu} + {(-ig)}^2 
\int\limits_x^y dt_{\nu} \int\limits_x^t dz_{\mu} A_{\mu}(z) A_{\nu}(t) + ...
\ee

The double brackets in (\ref{cum}) denote the irreducible cumulants, 
for example $\lll\lll AB\rrr\rrr = \lll AB\rrr -
\lll A\rrr\lll B\rrr$,
see \cite{clex} and \cite{dgs} for details.

Applying the nonabelian Stokes theorem \cite{st}
and cluster expansion property \cite{clex} one obtains:
$$
W(C) = \lll Tr\> Pexp\>(ig \int\limits_{C=\partial S} A_{\mu} dx^{\mu})\rrr =
 \lll Tr\> Pexp\>(i \int\limits_{S} G(x_i,x_0) d{\sigma}(x_i))\rrr =
$$
\be
= Tr\> Pexp\>(i \sum\limits_{n=1}^{\infty} D_{n} d^{n}{\sigma})
\label{clus}
\ee
where $D_{n}$ are defined in (\ref{cum}).
It may be proven, that the r.h.s. of (\ref{clus}) does not depend
on the reference point $x_{0}$ as it must be, if all $D_{n}$ are
taken into account.
Therefore there is a natural question, which $n$ are essential
in the expansion (\ref{clus}). One easy notes, that for small
contours, where the coupling constant is small, highest
cumulants which are by definition non-reducible Green's functions
are small too. 
Lattice calculations and some theoretical 
considerations demonstrate, that
the same conclusion is going to hold in the nonperturbative
regime \cite{dom} and only a few lowest cumulants
are important.
In particular, the lowest nontrivial bilocal cumulant
measured on the lattice in \cite{lat}
is believed to be dominant and it is
actually
the case in the so called Gaussian model of the gluodynamics
vacuum \cite{sss,dgs}.

Therefore it would be interesting to understand the perturbative 
behaviour of the lowest gauge-invariant functions (\ref{cum}).
First of all, one could compare the renormalization properties of 
(\ref{cum}) and (\ref{w}).
It will be seen below, that they are different, 
inspite of the similarities encoded in (\ref{clus}).
Second, such results may be useful for the lattice
calculations, where perturbative part of the cumulants is measured
at the small distances. It should also be noticed, that 
gluon propagator in the radial gauge is expressed as
an integral of the quantities like (\ref{d}) \cite{leup}.
So analysing perturbative expansion of (\ref{d}) we
get some information about the perturbative series in
Fock-Schwinger (radial) gauge.
   
In the present letter we are going to present leading perturbative 
contributions to
the simplest $n=2$
cumulant at the one-loop level:
\be
 D_{\mu\nu\rho\sigma}(x, y) =
Tr \lll gF_{\mu\nu}(x) U(x,0) U(0,y) gF_{\rho\sigma}(y) U(y,0) U(0,x)\rrr
\label{d}
\ee
The integration in any phase factor is along the straight line, connecting
its end points.

The detailed full calculation of this quantity
in the limiting case $y=0$
has been done recently
in \cite{ej}
by two methods -- direct calculation of the corresponding Feynman diagrams
and by methods of heavy quark effective theory.
If the reference point ($x_0 = 0$ in our case)  is chosen at
one of the correlator arguments (i.e. points where the
operators $F$ act are connected by the straight line), the quantity
(\ref{d}) depends on the only vector $z=(y-x_0) - (x-x_0)$
and two different tensor structures can be introduced:
\be {\dl}_{\mu\rho} {\dl}_{\nu\sigma} - {\dl}_{\nu\rho} {\dl}_{\mu\sigma} =
{\Delta}^{(1)}_{\mu\nu\rho\sigma} \ee
and
\be
{\Delta}^{(1)}_{\mu\nu\rho\sigma}
-\frac{d}{2}\left( \frac{z_{\mu}z_{\rho}}{z^2} {\dl}_{\nu\sigma} - \frac{z_{\nu}
z_{\rho}}{z^2} {\dl}_{\mu\sigma} + \frac{z_{\nu}z_{\sigma}}{z^2}
{\dl}_{\mu\rho} - \frac{z_{\mu}z_{\sigma}}{z^2}{\dl}_{\nu\rho} \right) =
 {\Delta}^{(2)}_{\mu\nu\rho\sigma}
\ee
Note, that
$$
{\Delta}^{(2)}_{\mu\nu\rho\sigma} {\dl}_{\mu\rho} {\dl}_{\nu\sigma} = 0
$$
therefore only the part proportional to ${\Delta}^{(1)}$ contributing
to the condensate $\lll {\alpha}_s F_{\mu\nu}F_{\mu\nu} \rrr $.
Two scalar coefficient functions
in front of 
${\Delta}^{(1)}$ and ${\Delta}^{(2)}$: 
\be
 D_{\mu\nu\rho\sigma}(x-y) =
 D(z)
{\Delta}^{(1)}_{\mu\nu\rho\sigma} + D_1(z) {\Delta}^{(2)}_{\mu\nu\rho\sigma}
\label{cr}
\ee
were actually calculated
in \cite{ej} and appeared to be nonzero both contrary
to the tree level results where only $D_1(z)$ presents.

Generally tensor structure of (\ref{d})
is more complicated and the results of \cite{ej}
should be recovered by taking the points $x, y$ and $x_0 = 0$
lying on a line, for example $x=-y = z/2$.
So our results can be used as an independent check
of \cite{ej}.

Let us briefly remind the situation in QED.
The abelian analog of (\ref{d}) was investigated in \cite{vain}.
Calculation there simplifies greatly since
phase factors are cancelled.
Representing renormalized photon propagator in Feynman gauge as
(we use Euclidean metric
throughout the paper):
\be \lll e A_{\mu}(x) e A_{\nu}(y)\rrr =
{\delta}_{\mu\nu}\>\frac{1}{4{\pi}^{2}}\>
\frac{d(z^2)}{z^2}
\label{aa}
\ee
where $z=y-x$ and
$d(z^2) = e^2(z^2)$ - the renormalized effective charge and
differentiating (\ref{aa}) with respect to $z$ one easily finds:
\be
\lll e F_{\mu\nu}(x) e F_{\rho\sigma}(y)\rrr =   \frac{1}{{\pi}^2}\> \left[
{\Delta}^{(2)}_{\mu\nu\rho\sigma}\left( \frac{d}{z^4} - \frac{d'}{z^2} +
\frac{d''}{2} \right) -
{\Delta}^{(1)}_{\mu\nu\rho\sigma}\>\frac{d''}{2}
\right] \ee
Here $z$ should be taken nonzero in order to avoid contact terms and
the derivative  $d' \equiv d e^2(z^2)/dz^2$.
Only the term proportional to $d''$ contributes to
$\lll F_{\mu\nu}F_{\mu\nu}\rrr$.
Using the definition from \cite{vain} for the
coordinate $\beta$-function :  $$ z^2 \frac{de^2}{dz^2} = \beta(e^2) $$
it is easy to see that
\be \lll e^2 F_{\mu\nu}(x)F_{\mu\nu}(y)\rrr = \frac{6}{{\pi}^2}\> 
\frac{1}{z^4}\> \left( 1 -
\frac{d\beta(e^2)}{d e^2} \right)\cdot \beta(e^2)
\ee
Taking $e^2(z^2) = e^2({\mu}^2) + b e^4({\mu}^2) ln [1/({\mu}z)^2] +
O(e^6)$
we find at the lowest nontrivial order:
\be
\lll e^2 F_{\mu\nu}(x)F_{\mu\nu}(y)\rrr = -\frac{6be^4({\mu}^2)}{{\pi}^2 z^4}
\ee

At the tree level in the $SU(N)$ Yang-Mills theory
 the only difference from abelian case in the common
colour factor $(N^2 - 1)/2$:
\be
D^{(0)}_{\mu\nu\rho\sigma} = \frac{N^2 - 1}{2} \>\frac{g^2}{\pi^2}\>
\frac{1}{z^4}\cdot{\Delta}^{(2)}_{\mu\nu\rho\sigma}
\label{tat}
\ee
Situation at the loop level is much more complicated
in the nonabelian case.
The main reason
is the presence of phase factors $U(x,y)$ in the
expression (\ref{d}).
Integrations over the straight lines,
connecting the points $x, y$ and $x_0 = 0$ lead to a new type of divergencies,
additional to the usual ultraviolet ones.
These divergencies are studied well in context of
renormalization of the Wilson loops and we will
explain below  the role they play in  renormalization of field correlators.

Let us proceed with the direct calculation of 
the leading $g^4/\epsilon$ terms in (\ref{d}).
The Feynman diagrams, contributing to the quantity under
consideration are presented in Fig.1 and Fig.2.
Naively only diagrams $1a,b,d$ and $2a,b,c,d,f,h$ give
the leading contribution.
Tree level contribution $D^{(0)}_{\mu\nu\rho\sigma}$
is defined in (\ref{tat}).
Using dimensional regularization $d=4-2\epsilon$
and keeping only $1/{\epsilon}$ terms we get in the Feynman gauge
(we consider gluodynamics but one can include dynamical quarks easily):
$$
1a) + 1b) = \frac{{\alpha}_s}{\pi} N\> \frac{5}{12} \left(\frac{1}{\epsilon}
\right) \cdot D^{(0)}_{\mu\nu\rho\sigma}
$$
$$
1d) = -\frac{{\alpha}_s}{\pi} N\> \frac34 \left(\frac{1}{\epsilon}
\right)\cdot D^{(0)}_{\mu\nu\rho\sigma}
$$
$$
2b)+ 2h) = \frac{{\alpha}_s}{\pi} N\>\left(\frac{1}{\epsilon}
\right)\cdot D^{(0)}_{\mu\nu\rho\sigma}
$$
$$
2d) = \frac{{\alpha}_s}{\pi} N \>\left(\frac{1}{\epsilon}
\right)\cdot\frac12 \left[ (\pi - \gamma)ctg(\gamma)\right]\cdot  
D^{(0)}_{\mu\nu\rho\sigma}
$$
In the last expression $\gamma$ is an angle between $x$ and $y$.
It may be seen that diagram $2a$ does not contain singular parts and
diagram $2f$ equals to zero.
The straightforward check shows the coincidence with the results 
for the leading logarithmic terms from \cite{ej}.
 Note,
that contrary to \cite{ej} we are working  in
the fundamental representation and our $D$ from
(\ref{d}) is in fact ${\cal D}/2$ from \cite{ej}.

All the above diagrams proportional to ${\Delta}^{(2)}$ and therefore
do not contribute into the condensate
$$
D_{\mu\nu\rho\sigma}(x,y) \cdot {\delta}_{\mu\rho} {\delta}_{\nu\sigma}
$$
it is fullfilled by the only $2c$, the result is:
$$
D_{\mu\nu\rho\sigma}(x,y) \cdot {\delta}_{\mu\rho} {\delta}_{\nu\sigma} =
$$
\be
= \frac{g^4}{16 {\pi}^4} N(N^2 - 1) \> \left(\frac{1}{\epsilon}
\right)\cdot \frac{1}{z^4} \>\left( \left[ 1 - 4 \frac{{(xz)}^2}{x^2
z^2} \right] +
 \left[ 1 - 4 \frac{{(yz)}^2}{y^2
z^2} \right] \right)
\label{ll}
\ee
We note the nontrivial dependence on the angles between 
$x$, $y$ and $z=y-x$, expressed via   
scalar products $(xz)$ and $(yz)$.
In particular case,
if the special position of the points $x$, $y$ and $0$
is chosen the leading logarithmic contribution to 
the condensate is absent and only finite renormalization 
is needed.
Namely, it is the case if the points
$x$, $y$ and $0$ form the equal-side triangle
of arbitrary size.  
This circumstance might be useful in the lattice
calculations for fixing the normalization scale
of the leading logarithms.

We see that the one-loop contribution to the field 
correlators cannot be entirely explained by charge 
renormalization. It is clearly indicated  also in \cite{ej}
where besides terms proportianal to the first coefficient of 
the $\beta$-function
there are additional terms in $D_{1}$ and $D$.

To proceed, let us compare 
the renormalization procedure for the Wilson loops
 (see \cite{sato} for
details) and correlators.
The former depends on whether the corresponding Wilson contour $C$ is
smooth or not despite the final results look in some sense similar.
In the former case the 
average is renormalizable in the following sense:
\be W(C) = Z\cdot
W_{ren}(C) \label{ren}
\ee
where infinite $Z$-factor contains only linear
divergencies arising from the integrations over the contour:
\be
Z\sim {\exp}\left({c\>\int\limits_{C} dx_{\mu}\int\limits_{C} 
dy_{\nu} D_{\mu\nu}(x-y)}\right)
 \sim {\exp}\left({c\> \frac{L}{a}}\right)
\ee
where $L$ is the length of the contour $a$ -- ultraviolet cutoff and $c$ --
numerical constant not important for us here.
At the same time logarithmic divergencies are absorbed in a usual
way, so that
renormalized Wilson average $W_{ren}$ in the Eq. (\ref{ren}) is functional
of the contour $C$ and function of the renormalized coupling constant
$g_{ren}(\mu)$ defined on the corresponding dynamical scale $\mu$.

It is worth noting that a single diagram in the perturbative 
expansion of the r.h.s. of (\ref{ren}) typically gives contributions to 
both $Z$-factor and $W_{ren}(C)$, i.e. contain linearly and
logarithmically
divergent parts. Hence the factorization in the r.h.s. of (\ref{ren})
has double-faced nature - it is factorization of the diagrams together
with the separation of contributions, coming from dangerous integration 
regions (i.e. linearly divergent terms) for each diagram itself.

As a result contour divergencies can be separated as a common 
$Z$-factor, which physically renormalizes the bare mass of the  
test particle, moving along the Wilson contour.
This additional infinite constant in the exponent reflects 
the impossibility to observe the Wilson
loop itself. At the same time if one extracts the physical 
quantities from the Wilson loop average 
then only logarithmic renormalization
cause observable effect. 
 
If the contour possesses cusps (and selfintersections) situation
becomes more complicated \cite{sato}. Namely, each cusp leads to its
own local $Z$ -- factor, depending on the cusp angle (but not on another 
characteristics of the contour). The Eq.(\ref{ren}) for the loop with 
$K$ cusps having angles ${\gamma}_i$ is to be modified
in the following way:
\be W(C) = \prod\limits_{i=1}^{K} Z_i\>({\gamma}_i)\cdot 
W_{ren}(C) \label{ren1}
\ee

The mixing between linear and logarithmic divergencies, mentioned
above has some specific features for the loops with cusps.
Namely, each cusp in addition to the linearly divergent factors, removed 
by the test particle mass renormalization introduces logarithmically  
divergent terms which have nothing to do with the coupling constant 
renormalization and must be subtracted by their own 
$Z({\gamma}_i)$--factors. There are two types of terms of this 
kind, which do 
depend on the cusp angle 
and which do not \cite{sato}. The latter plays important role in
the perturbative expansion of the correlators. 
To see this, note, that the correlators can be
easily obtained from the Wilson loop functional.
Indeed, differentiating (\ref{w}) with respect to
${\delta}{\sigma}_{\mu\nu}(x_i)$
and assuming that the contour $C$ connects the points $u, v$ with the 
reference point $x_0$ along the straight lines, one finds for the bilocal
correlator:
\be
\frac{1}{N_c} Tr g^2 \lll\lll F_{\mu\nu}(u,x_0) F_{\rho\sigma}(v,x_0)
\rrr\rrr =
- \frac {{\delta}^2 W(\tilde C)}{\delta {\sigma}_{\mu\nu}(u) \delta 
{\sigma}_{\rho\sigma}(v)}
\label{qu}
\ee
and analogous expressions for higher cumulants. Note, that area
of the minimal surface bound by the contour $\tilde C$
from (\ref{qu}) is equal to zero and this contour generally has 
at least two cusps (typically even four, if the reference point
$x_0$ does not coincide with one of the correlator arguments). 

Comparing Eqs. (\ref{ren1}) and (\ref{qu}) one concludes, that 
the derivatives in the r.h.s. of (\ref{qu}) must act on renormalized
part of the Wilson loop as well as on the $Z$ -- factors.
As a result mixing between different terms appears 
and contour divergencies 
cannot be separated in some
simple way as they can in the case of loops, so the perturbative
behaviour of the correlators is controlled by both ultraviolet 
and contour logarithmic divergencies.

Let us do the final remark.
It may be easily shown, that taking into account only the bilocal
correlator
in the cumulant expansion (\ref{clus})
one has perturbative contributions to the
string tension at the one loop level since the kroneker part of (\ref{cr}) 
does not equal to zero.
But area law is perturbatively impossible.
The problem is cured by the contributions from the higher cumulants, 
in particular triple correlator $\lll FFF\rrr$ also contain such
kroneker parts being considered at the same ($g^4$ in our case)
order of perturbation theory \cite{ssd}. As a result string tension 
is equal
to zero in the field-strength formulation as it must be at any given
order of perturbation theory if all cumulants contributing
to string tension at this order are taken into account.
Note in this respect that there are no terms proportional to the 
$\beta$-function coefficients in the function $D$ \cite{ej} - 
it is a direct consequence of the fact that
nonzero $D$ has no physical sense in perturbation theory and 
all contributions into it must be cancelled by the next-order 
terms.

\bigskip

{\bf Acknowledgements}

The work was supported by RFFI grants 96-02-19184 and RFFI-DFG 
96-02-00088G. The author is grateful to the theory group
of Heidelberg University where the part of the work was
done for kind hospitality. Discussions with H.G.Dosch,
M.Eidem\"uller, M.Jamin are gratefully acknowledged.
The author would also like to thank Yu.A.Simonov for useful 
discussions and critical remarks and A.Vairo for pointing 
authors attention to the Ref.[6].

\newpage

\begin{figure}[thb]
\epsffile{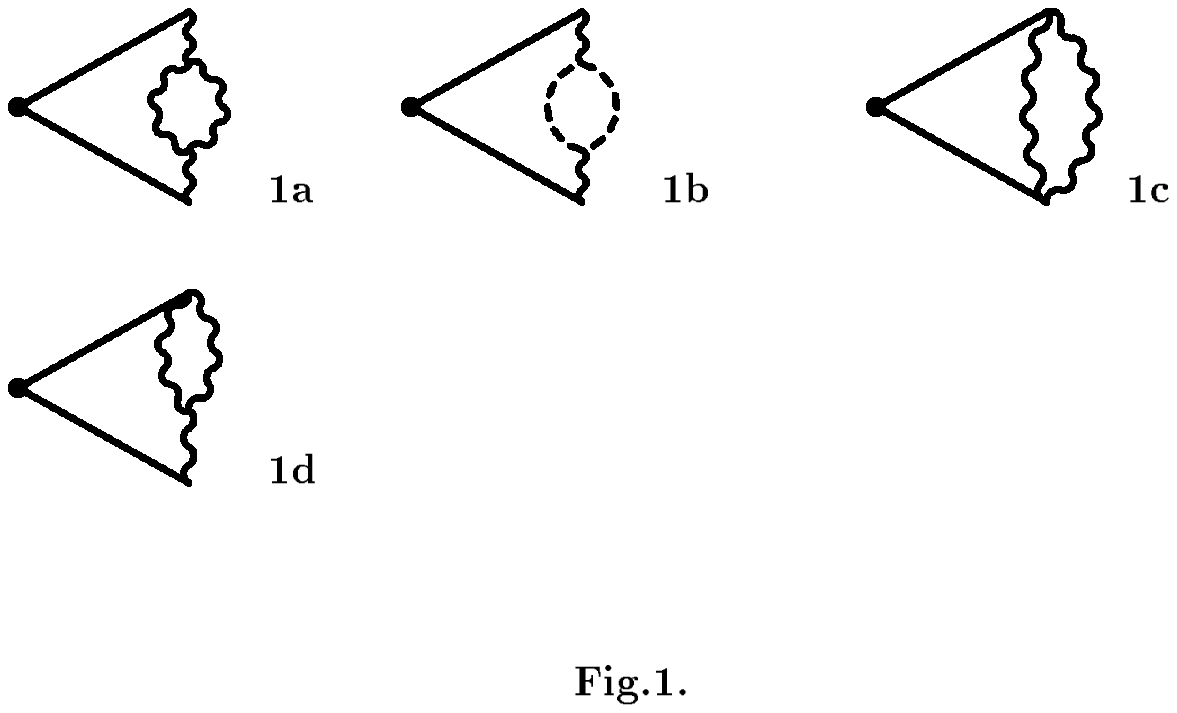} 
\end{figure}
\newpage
\begin{figure}[thb]
\epsffile{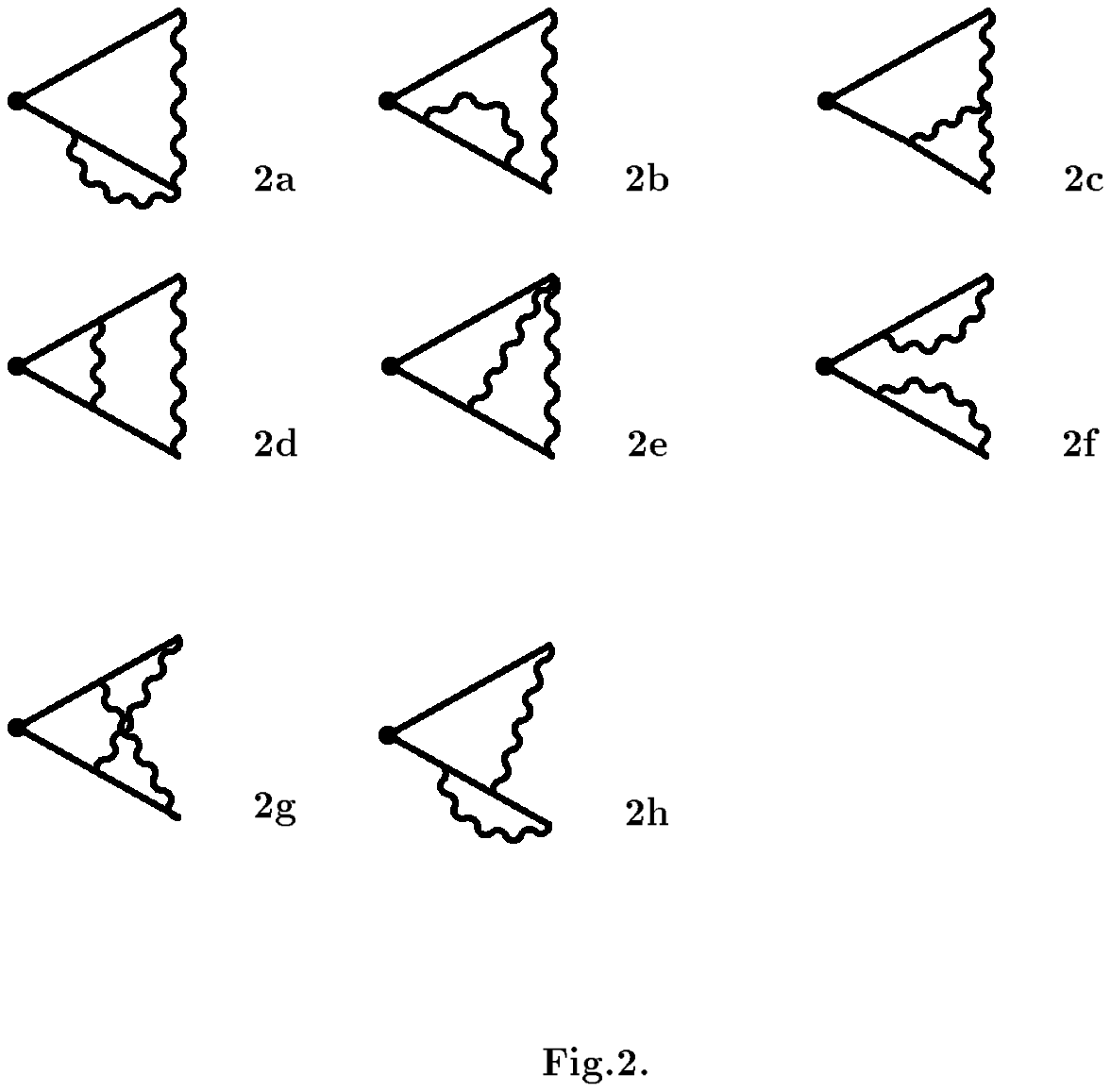} 
\end{figure}

\end{document}